# Automated Imaging of the Annihilation of a Transverse Domain Wall in Patterned Magnetic Thin Films

Charudatta Phatak[1,2*], John Fullerton[1], and Hanu Arava[1]

[1]Materials Science Division, Argonne National Laboratory, IL, USA.

[2]Department of Materials Science and Engineering, Northwestern University, IL, USA.

*cd@anl.gov

*Imaging the magnetic domain wall behavior in patterned thin films under external stimuli can enable understanding the underlying energy landscape as well as the role of local microstructure and defects. We present an automated workflow for in-situ Lorentz transmission electron microscopy to image magnetic domain walls at the nanometer length scale and at a time resolution in the sub-millisecond regime—the latter of which is limited by the speed of the available camera. Our workflow is modular and can be broadly applied to various types of in-situ experiments, taking us a step closer to the future of autonomous imaging of nanomagnetic films with electron microscopy. Using our workflow, we show the transformation of a transverse domain wall with sub-millisecond time resolution under the application of an in-situ transverse magnetic field, a study of whose dynamics are essential in the design of future domain wall mediated spintronic device applications.*

Understanding of nanoscale magnetic spin textures and their behavior is of critical importance for both their fundamental properties as well as for technological applications in spintronics and novel energy efficient information processing systems[1-11]. From a fundamental perspective, there is ongoing research on the formation of novel types of domain walls (DWs) as well as topological spin textures within them such as DW Skyrmions, and Bimerons[12]. Microstructural and patterned features such as notches have been used for controlling the pinning and depinning fields for DW motion[13]. There is still a need to understand the motion of DWs and related transformations to gain insight into the underlying physics and materials effects. One technique to achieve this is by Lorentz transmission electron microscopy (LTEM), which is an increasingly used powerful technique that enables real-space imaging with high spatial resolution of both the magnetic domain structure and sample microstructure[14,15]. LTEM has been used for direct visualization of Skyrmions[16,17], domain walls[18-20], as well as most recently magnetic spin texture at the atomic scale in antiferromagnets[21]. Furthermore, LTEM can be used with a wide range of *in-situ* experiments using specialized sample holders for electrical biasing[22,23], magnetic field application[24,25], temperature control[26] while imaging. Recently there is also progress towards improving the time resolution in LTEM imaging through development of fast cameras as well as stroboscopic ultrafast techniques[27-29].

Despite its capabilities, conducting *in-situ* LTEM experiments poses significant challenges. The intricacies of manually controlling imaging conditions, such as beam alignment, and defocus, alongside the precise application of external stimuli, make these experiments time-consuming and susceptible to human error. Additionally, the dynamic nature of magnetic processes demands rapid data acquisition and real-time analysis, which are difficult to achieve with manual operation. To address these challenges, automated control systems are becoming increasingly important. Automation can enhance the efficiency and reproducibility of experiments, reduce potential human-induced variability, and enable the exploration of a larger experimental parameter space.

Furthermore, automated systems lay the groundwork for autonomous experimentation, where machine learning algorithms can drive the experimental workflow based on real-time data analysis[30]. Electron microscopy is experiencing such developments of automated and autonomous experiments including segmentation of images[31,32], determination of atomic structure and correlation with spectroscopic information[33,34], and *in-situ* forecasting[35].

In this work, we have developed an automated control workflow for *in-situ* LTEM specifically tailored to image magnetic domain wall motion. By integrating and developing an open-source Python based scripting capabilities for microscope control, camera control, and external device control, we have established a system that can manage imaging parameters, beam alignment, and application of external stimuli in a programmable manner. We demonstrate the effectiveness of our workflow by imaging the magnetic domain wall motion in a Permalloy ($Ni_{80}Fe_{20}$) strip under applied magnetic fields. Utilizing our automated workflow, we are able to achieve sub-millisecond time resolution to capture the magnetic domain wall dynamics.

***Development of the automated workflow for in-situ imaging in LTEM***: The magnetization reversal studies were performed on Permalloy thin film of 20 nm thickness that was deposited directly on Si/SiN TEM grids. The thin film was patterned using direct milling using the Ga+ ions in a focused ion beam instrument. A JEOL 2100F equipped with a Lorentz pole piece and an aberration corrector was used for the LTEM imaging of the domain walls in the Fresnel mode. The *in-situ* magnetic field was applied using a specialized holder from Hummingbird Scientific which is capable of applying uniform in-plane magnetic fields up to ±750 Oe. The field is applied by connecting the holder to a source meter and applying current through the coils in the holder. The entire experimental setup is shown schematically in Fig. 1a in which (i)an open-source python library, PyMeasure[36], is used to interface with and control the source meter ; and (ii) a python script in Digital Micrography is used to acquire images and add additional metadata to the images from the microscope and the source meter. In Fig. 1b we show the under-focused LTEM image of the patterned Permalloy stripe in which the bright and dark magnetic domain walls are indicated by the white and black arrows. The relation between the various coordinate frames associated with the CCD camera or the image, the beam shift coils of the microscope, and the applied field from the in-situ holder are found in Fig. 1c. When a current step is applied via the source meter to the *in-situ* magnetizing holder to generate a magnetic field in the sample plane, it results in a beam shift and a consequent image shift in a direction perpendicular to the applied field. During *in-situ* magnetizing experiments, this is often the hurdle as the region of interest (ROI) in the sample shifts or goes out of view making the analysis on-the-fly difficult as well as often the fast magnetic domain wall transitions are not captured. In order to correct for these shifts, and perform the magnetizing experiments in an automated manner, we developed a workflow which pre-shifts the image using the microscope image shift coils by a calibrated amount as per the strength of the magnetic field to be applied. The field step is then applied such that the resulting image shift due to beam deflection is captured on the CCD, and the ROI stays within the field of view. Our workflow would determine the estimated image shift for a given applied current for the field step and then convert that into image shift coils current that needs to be applied.

In order to demonstrate the viability of our workflow for automated acquisition, we applied a large positive and negative field to the sample and acquired the images of the sample. The under-focused LTEM image of the sample is shown in Fig. 3a. The LTEM images acquired at a negative applied field of 67.5 Oe and positive 45 Oe without any corrections are found in Fig. 3b. It can be seen that the ROI

(red arrow) in Fig. 3a is no longer visible in the negative field image, and only partially visible in the positive field image. This would make understanding the domain behavior under applied fields very difficult. Additionally events such as domain wall collapse or merging of domain walls under applied field would not be able to be imaged. The results from our automated acquisition workflow for the same values of applied fields are found in Fig. 3c, where the ROI of interest (indicated by red arrow) is now in the field of view. There is a slight movement of the images, but this is understandable due to the hysteresis in the electromagnetic coils both for the magnetizing holder as well as the microscope image shift coils.

Next, we applied the automated acquisition workflow to study the domain wall behavior in the patterned Permalloy stripe using the CCD camera operating in the standard mode. The applied magnetic field steadily increased from 0 Oe to 39 Oe corresponding to applied current of 14 mA in steps of 2.8 Oe (1 mA) and then decreased back to zero. The field was applied in a direction perpendicular to the length of the stripe as indicated by the red arrow in Fig. 5. In the top row in Fig. 5, you can find three sequential images corresponding to decreasing applied magnetic field, in which we observe the formation of closure-type domains. The magnetization directions in the domains are indicated by large white arrows. It should be noted that these images were automatically acquired through the continuous *in-situ* acquisition script. The bottom row of Fig. 5 is a sequence of images acquired at the constant field of 17.5 Oe, using the *in-situ* mode of our CCD camera at a time resolution of 5 ms. We observed the merging and collapse of domain walls, in which the two vertical domain walls (black and white) merge and the domains are rearranged to minimize energy. The resulting domain configuration is then dictated by the strong shape anisotropy of the patterned stripe giving rise to the staircase like magnetization pattern as indicated by white arrows in Fig. 5 at t = 10.665 s.

This time resolved LTEM imaging of magnetic domain wall behavior was enabled due to the automated acquisition which corrects the image shift due to applied magnetic fields thus maintaining the ROI in the CCD field of view. This increases the possibility of capturing rare events such as the domain wall merging observed in the Permalloy strip. Furthermore, the interaction of domain walls with defects and the microstructure of the sample can be observed during the in-situ acquisition process which was only possible previously through post-acquisition image alignment and data analysis.

***Fast imaging of transverse domain wall reversal in a transverse magnetic field***: Next, we made use of our automated workflow in conjunction with the IDES/Relativity electrostatic sub-framing option[37] in the LTEM to image dynamics of a transverse domain wall. Such a sub-framing option enables us to access timescales previously not accessible using the standard LTEM CCD camera. Now, we are able to achieve a time resolution in the sub-millisecond regime with the possibility of reaching sub-microsecond for particular needs. We carefully selected a region of interest that is extremely relevant to the design of domain wall devices[3,38,39], which contains a prototypical transverse domain wall, circled in red in Fig. 4f. An appropriate sub-framing size of 7×7 was selected, which provided us with a temporal resolution of ~0.8 ms per captured frame. The region presents us with a transverse domain wall to study field dependent domain wall dynamics for which experimental evidence has been sparse. A micromagnetic simulation of the observed structure is provided in Figs. 5a-c, all correspond to increasing time and thus sequence of domain wall evolution. We have also included the corresponding simulated images for what to expect in the LTEM (Figs. 5d-f). By correlating the Mumax3[40] micromagnetic simulations with the experiment (Figs. 5g-h), we are able to observe the annihilation of a transverse domain wall under a transverse field and gain deeper insight into the underlying physics governing the behavior..

The transverse field was applied with a step size of 2.8 Oe, the direction applied field is depicted in Fig. 5. We do not observe any changes to the domain wall structure until up to a field of ~19.13 Oe (Fig. 4f), after which an antivortex is nucleated at the top of the transverse domain wall (19.12 Oe), which is then subsequently annihilated (21 Oe) by the previously present clock-wise (CW) vortex wall—experimentally, this was found to occur within a timeframe of 0.8 milliseconds. The domain wall is later transformed into a relatively stable structure at 23 Oe and remains the same throughout the application of the field protocol—which was terminated at ~100 Oe. Our micromagnetic simulations of a similar structure do indeed confirm the emission of an antivortex domain wall from the top of the transverse domain, which agrees with existing literature[20,41-43]. Upon carefully observing the relevant energetics of the simulations, we find that such an emission of an antivortex domain wall is indeed energetically favorable, which then proceeds to annihilate with the previously present vortex domain wall leading to the further lowering of energy (Fig. 5j). A topological argument can also be made here, whereby the domain wall reversal will conserve the net winding number in the structure. We observe such a conservation of winding number in our micromagnetic simulations (Figs. 5k-m), here the edge of the transverse domain wall with an initial winding number of $+\frac{1}{2}$ transforms to a $-\frac{1}{2}$ upon the completion of domain wall reversal. It should be noted that globally, the transverse domain wall retains a winding number of 0 upon domain wall reversal. By combining the automated workflow with the IDES/Relativity measurements, we provided real-space imaging evidence for the nucleation and the subsequent annihilation of an antivortex wall under the application of a transverse magnetic field to a transverse domain wall.

Our automated workflow is the first step towards building an autonomous LTEM platform, where based on the live image analysis, decisions can be taken, and magnetic domains can be analyzed. For example, in our previous work[44], we have developed a convolution neural network (CNN) based method for automated identification of Neel-type Skyrmions in LTEM images. The trained CNN can be coupled with our data acquisition pipeline to control the applied field or electric current that can affect the motion of the Skyrmions. This can further be applied to capture rare events such as changes in magnetic domain topology which can change from non-trivial to trivial state under applied magnetic fields. Currently all the automation is implemented within the Digital Micrograph Python scripting environment and the script is available via GitHub. We are currently expanding the workflow to various in-situ LTEM experiments such as in-situ Hall transport measurements where magnetic domain configuration can be directly correlated with measured Hall resistance.

**Author Contributions**

All authors contributed equally to this work.

**Author Declarations**

The authors have no conflicts of disclose.

**Acknowledgements**

This work was funded by the US Department of Energy, Office of Science, Office of Basic Energy Sciences, Materials Science and Engineering Division. Work performed at the Center for Nanoscale Materials, a U.S. Department of Energy, Office of Science User Facility, was supported by the U.S. DOE, Office of Basic Energy Sciences, under contract no.DE-AC0206CH11357.

**Data Availability Statement**

The data that supports the findings of this study are openly available in Zenodo at http://doi.org/10.5281/zenodo.14994083.

**Figures**

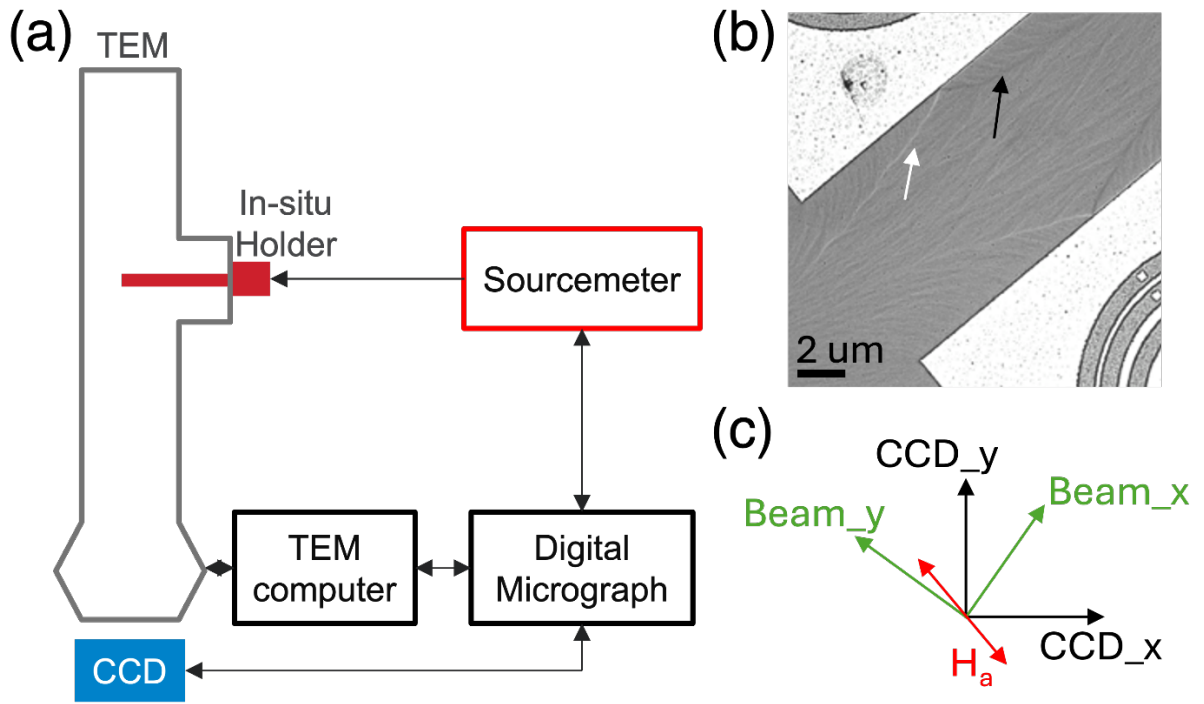

FIG. 1. (a) Schematic of the experimental setup for automated control of in-situ LTEM. The python scripts are run in the Digital Micrograph environment; (b) LTEM image (out-of-focus) of patterned NiFe thin film showing the locations of magnetic domain walls (bright and dark); (c) the x – y coordinates and their orientation relation between the OneView camera (CCD), the beam shift coils, and the applied magnetic field ($H_a$).

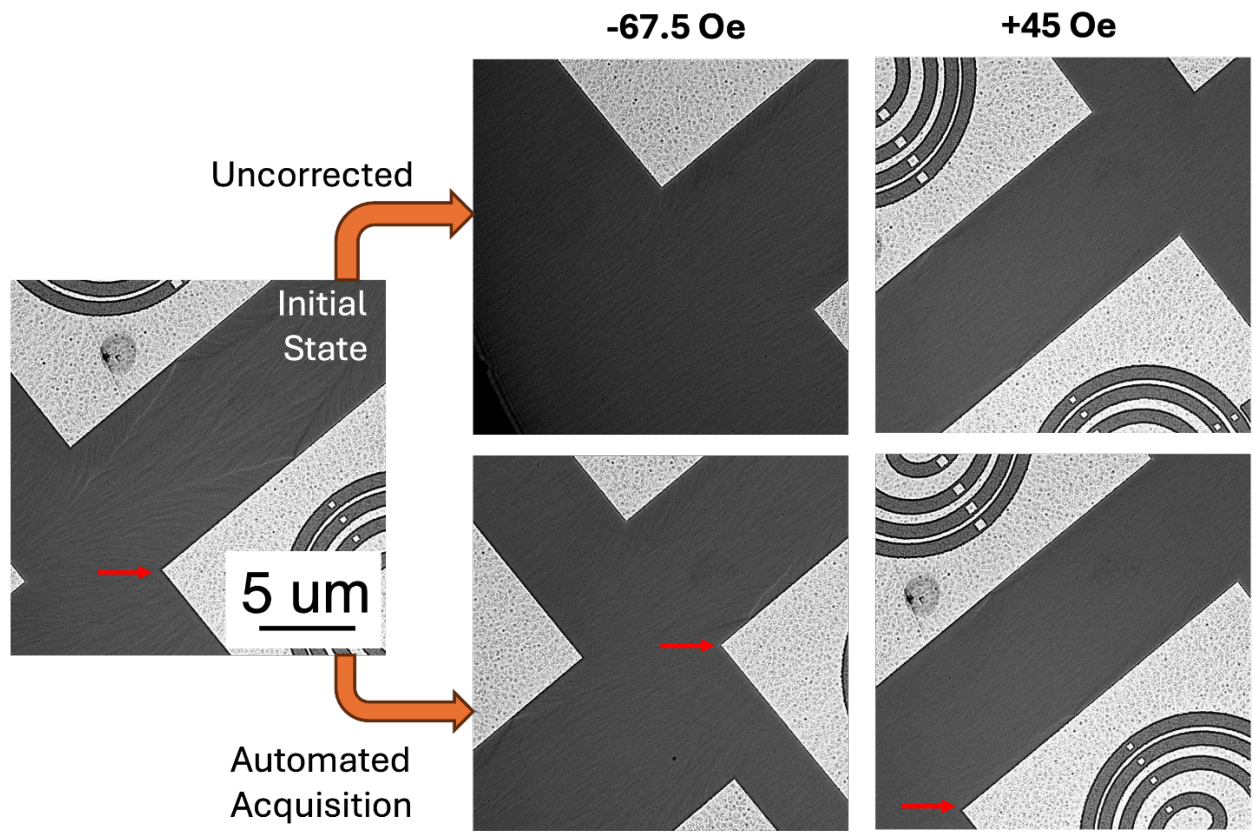

FIG. 3. (a) Under focused LTEM image showing the initial state of the sample. (b) Images acquired under applied field of −67.5 Oe and +45 Oe without any correction, and (c) images acquired for the same field values using automated acquisition.

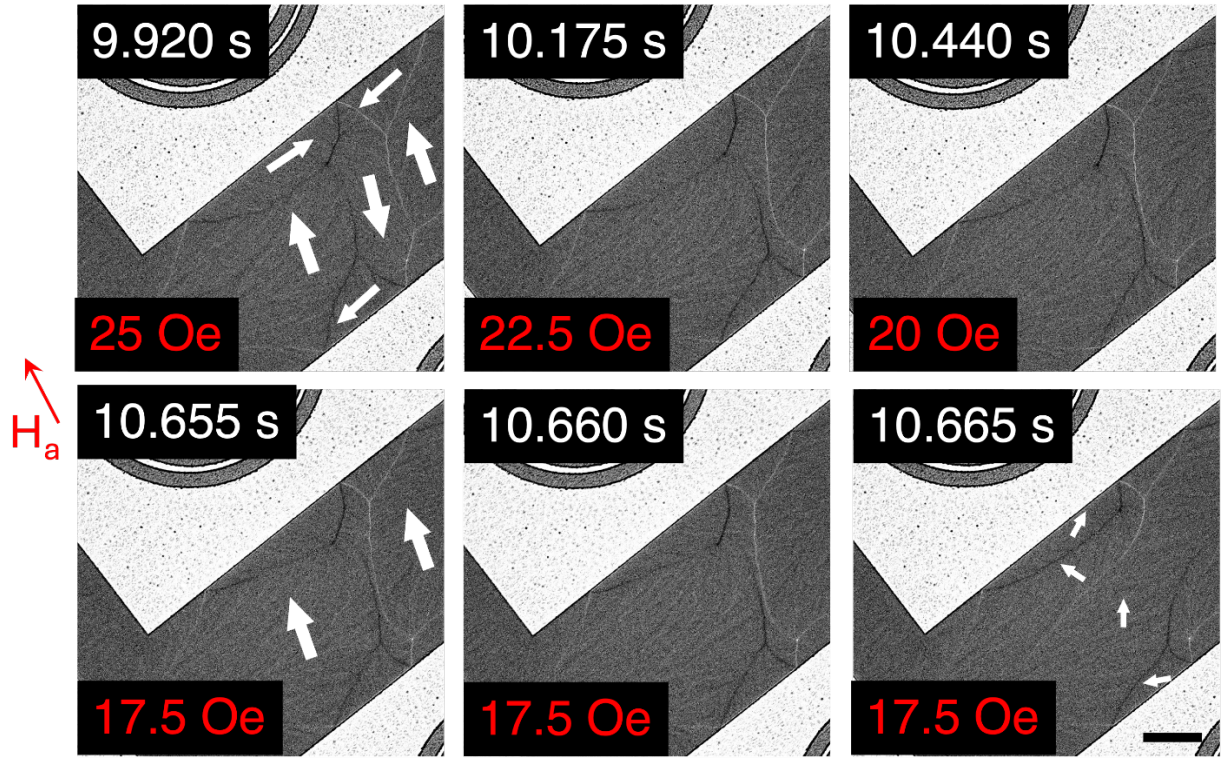

FIG. 4. Series of under-focused LTEM images acquired during the in-situ magnetization reversal process of the patterned Permalloy strip.

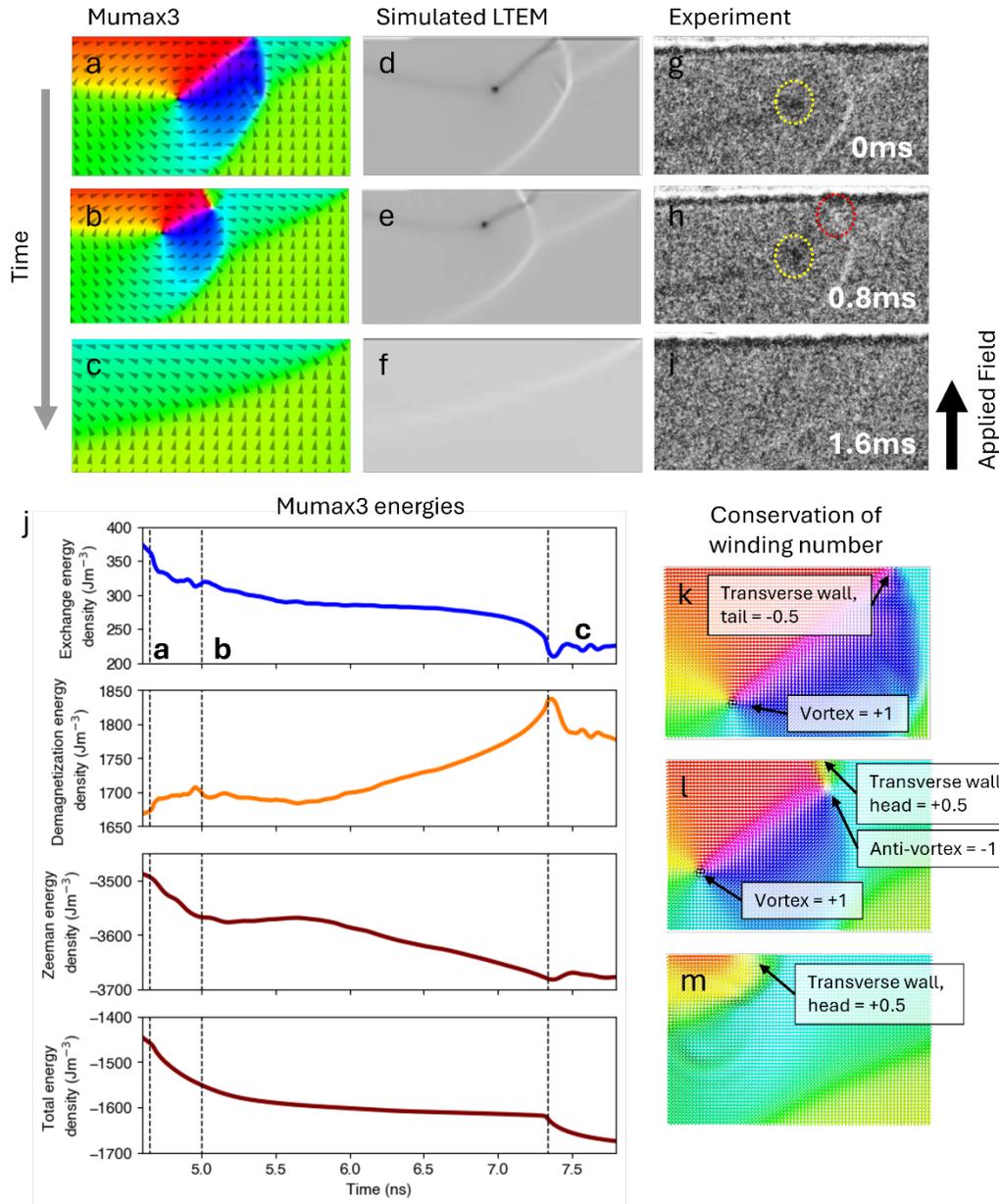

FIG. 5. (a-c) Micromagnetic simulations of the structure found in the IDES/Relativity experiment. (d-f) Simulated LTEM images for the micromagnetic simulations in a-c. (g-i) The experimentally observed sub-millisecond dynamics of domain wall reversal under a transverse magnetic field. (j) The energetics found in micromagnetic simulations, labeled with relevant events corresponding to the images in a-c. (k-m) Local conservation of winding number found in the micromagnetic simulations, which is observed in the experiments (g-i)